\documentclass[
  reprint,
  amsmath,amssymb,
  aps,
  prl,
  floatfix,
]{revtex4-2}
\usepackage{graphicx}
\usepackage{dcolumn}
\usepackage{bm}

\usepackage{braket}
\usepackage{graphicx}  
\usepackage{siunitx}

\begin{document}


\title{Spectroscopy of $^4$He at 0.25 ppt Uncertainty and Improved Alpha–Helion Charge-Radius Difference Determination}


\author{K. Steinebach}
\author{J.C.J. Koelemeij}
\author{H.L. Bethlem}%
\author{K.S.E. Eikema} \email{k.s.e.eikema@vu.nl}%

\affiliation{
Department of Physics and Astronomy, LaserLaB, Vrije Universiteit Amsterdam, De Boelelaan 1100, 1081 HZ Amsterdam, The Netherlands
}

\date{\today}
\begin{abstract}
High-precision spectroscopy of simple atomic systems can be used to advance the theory of atomic energy levels but can also serve as a sensitive probe of nuclear charge radii. For this last purpose, we report an improved measurement of the  $2\,^3{S}_1 \to 2\,^1{S}_0$  transition frequency in $^4$He with 48~Hz uncertainty (0.25 ppt), using a Bose–Einstein condensed sample confined in a magic-wavelength optical dipole trap. A systematic Doppler shift from condensate motion is suppressed by time-resolved ion detection, and the transition frequency is calibrated via a White Rabbit link to a remote active hydrogen maser clock. Combined with previous $^3$He measurements and improved theory, we obtain the most precise determination to date of the charge-radius difference between the helion and alpha particle $(r_{h}^2 -r_{\alpha}^2)$ of $1.0676(10)\text{~fm}^2$. This is consistent with other recent determinations and confirms that the current discrepancy between QED theory and experimentally observed ionization energies of excited states in helium is not apparent in the isotope shift.
\end{abstract}

\maketitle

Precision measurements of the energy structure of simple atomic and molecular systems have provided some of the most stringent tests of quantum electrodynamics (QED) and, in general, of the standard model; see e.g.~\cite{Tiesinga2023,morgner_stringent_2023,ADKINS2022,Delaunay2023}. The fundamental constants required for theory calculations can be obtained by combining theory with high-precision experimental data~\cite{Tiesinga2021,Delaunay2023}. By comparing the results based on different systems, a valuable consistency check can be done in the comparison between experiment and theory. In this way, spectroscopy of electronic \cite{Beyer2017,Fleurbaey2018, Bezginov2019} and muonic hydrogen~\cite{Pohl2010} has led to adjustments of the Rydberg constant and the proton radius~\cite{Tiesinga2021}.

Helium, as the simplest multi-electron system after hydrogen, serves as an important platform for testing QED and two-electron effects. Combined with the experimental benefits of its laser-coolable and efficiently detectable $2\,^3{S}_1$ state, helium provides a unique platform for testing fundamental physics~\cite{vanderWerf2025, Henson2022}.

QED calculations of triplet helium (both electron spins parallel) have advanced to a level that includes corrections up to $m\alpha^7$ \cite{Pachucki2021}. This means that finite-size effects from the nucleus start to play an important role. However, significant discrepancies exist between these theoretical calculations and the respective measurements of the $2\,^3S_1\!\rightarrow\!3\,^3D_1$ transition \cite{Dorrer1997}, $2\,^3P_0\!\rightarrow\!3\,^3D_1$ transition \cite{Luo2016}, and $2\,^3S_1$  ionization energy \cite{Clausen2021, Clausen2025}. So far, it is unknown where these discrepancies stem from. This makes it impossible to determine the absolute size of the helium nucleus directly from spectroscopic measurements. 

Instead, the isotope shift on a transition in $^4\text{He}$ and $^3\text{He}$ can be measured. The cancellation in the isotope shift of mass-independent terms allows subkilohertz accuracy of the theoretical calculations, so that the differential nuclear charge radius squared $r_{h}^2 -r_{\alpha}^2$ between the alpha and helion particles can be determined from it~\cite{Pachucki2017}. Comparing measurements of $r_{h}^2 -r_{\alpha}^2$ from different systems provides a broad consistency check, testing not only the various experiments but also the different calculations required to extract $r_{h}^2 -r_{\alpha}^2$. 

This is illustrated by the latest developments: recently, in \cite{Wen2025} a large discrepancy \cite{Zheng2017} between various measurements \cite{Canciopastor2004, Canciopastor2012, Shiner1995} of  $r_{h}^2 -r_{\alpha}^2$ from the $2\,^3S_1\rightarrow2\,^3P$ was resolved by identifying a systematic postselection shift. Also, the measurement from the \(2\,^3S_1 \to 2\,^1S_0\) was improved \cite{vanderWerf2025}, and the  previous difference  of \cite{vanRooij2011} with the \(2\,^3S_1 \to 2\,^3P\) in helium was explained by a systematic shift due to the interplay between the Fermi gas and the optical trapping potential. A comparison of $r_{h}^2 -r_{\alpha}^2$ obtained with electronic (regular) helium with a measurement from muonic helium ions revealed a surprising 3.6$\sigma$ difference \cite{Schuhmann2025}. This inspired more advanced calculations \cite{Qi2025, Pachucki2024} of the hyperfine mixing in $^3\text{He}$ revealing a shift of $-1.7$~kHz in the isotope shift calculation for the $2\,^3S_1 \rightarrow 2\,^1S_0$ for a point-like nucleus~\cite{Pachucki2024}, reducing the difference of $r_{h}^2 -r_{\alpha}^2$ compared to muonic helium ions to 1.2$\sigma$\cite{vanderWerf2025}. Interestingly, $r_{h}^2 -r_{\alpha}^2$ has also been determined  from the isotope shift of the ionization energies of the $2\,^3S_1$ state \cite{ClausenMerkt2025}. Although there is a discrepancy between theory and the calculated ionization energy, the determined $r_{h}^2 -r_{\alpha}^2$ agrees with measurements from both muonic and electronic helium, suggesting that the discrepancy is not caused by finite size effects. 

In this work, we present the most precise frequency measurement in helium on the $2\,^3S_1 \to2\,^1S_0$ transition in $^4\text{He}$, leading to an improved determination of the charge radius difference between the helion and the alpha particle. The transition is observed in a $^4\text{He}$ Bose-Einstein Condensate (BEC), confined in a magic-wavelength optical dipole trap. Calibration of the transition frequency is performed via a White Rabbit fiber link referenced to a remote active hydrogen maser. We developed a method to observe the motion of the BEC in the optical trap and methods to reduce this motion and its effects on the calibration. 
In addition, we determined the singlet-triplet scattering length and the magic wavelength more precisely.

\begin{figure}[b]
\includegraphics[width = 0.9\linewidth]{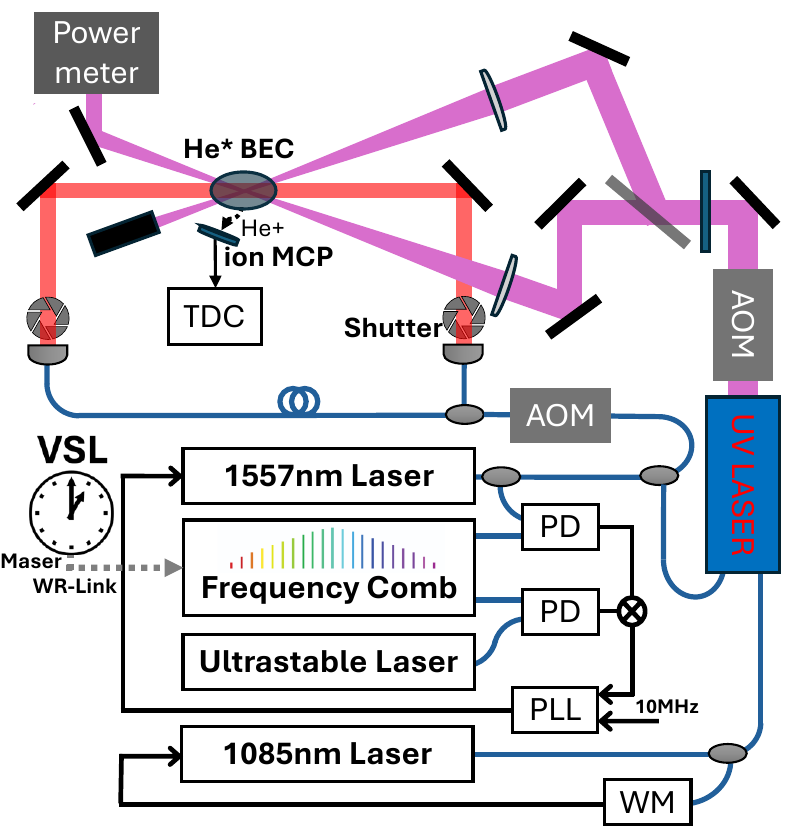}
\caption{Overview of the optical dipole trap and frequency metrology. Using a phase-locked loop (PLL), the 1557\,nm spectroscopy laser is locked to an ultrastable laser via an optical frequency comb, by mixing the beat note from a photodiode (PD). The OFC is referenced through a White Rabbit (WR) link to an active hydrogen maser at VSL. Part of the 1557\,nm light is used together with a separate 1085\,nm laser, locked to a wavemeter (WM), to generate the UV laser light. The UV laser light is sent through an AOM, and subsequently split into two beams of orthogonal linear polarization that are focused to form the two arms of the optical dipole trap. The BEC is trapped at the crossing point of the two beams. By using shutters, we excite the BEC alternately from opposite directions. The He$_{(2)}^+$ ions created by the excitation are detected using an MCP, and the arrival times are registered using a time-to-digital converter (TDC).}
\label{fig:experimental_setup}
\end{figure}

The starting point for our spectroscopy measurements is a BEC of $^4\text{He}$ in the metastable $2\,^3{S}_1$ state trapped in an optical dipole trap near the magic wavelength  of 319.816~nm. A schematic overview of our experimental setup is shown in Figure~\ref{fig:experimental_setup}. Details on how we produce quantum degenerate gasses of He* in our experiment can be found in \cite{Rengelink2018, vanderWerf2025, Jannin2022}.

The dipole trap light is generated by frequency-doubling 640~nm light, produced via sum-frequency generation of amplified 1557~nm spectroscopy light and a separate amplified 1085~nm laser, whose wavelength is stabilized to a precision wavemeter (High Finesse WSU-30) to set the trap wavelength. The power of the dipole trap laser is controlled using an acousto-optic modulator (AOM). This enables control over the number of atoms and the chemical potential of the BEC by ramping down the trap power to expel atoms, followed by ramping it back up to obtain the desired trap depth.

The spectroscopy laser (bandwidth $\Delta\nu\approx400\text{ Hz}$) is transfer-locked to an ultrastable laser via an optical frequency comb \cite{Telle_Lipphardt_Stenger_2002}. The frequency comb and counters used for metrology are referenced via a 110~km long White Rabbit fiber link to an active hydrogen maser located at VSL, the national metrology institute of the Netherlands. This link reaches a stability in the low $10^{-14}$ after $1000~\mathrm{s}$ of averaging, with a noise floor of $5 \times 10^{-15}$.

The atoms in the BEC are all prepared from the start in one of the two spin-stretched states to suppress collision-induced Penning ionization. We then expose the BEC to spectroscopy light for 100~ms. When a triplet atom is excited to a singlet state, it collides with one of the many surrounding triplet atoms and Penning ionizes, producing a He$^+$ or He$^+_{2}$ ion. These ions are detected using a negatively biased microchannel plate (MCP), and the detection times of these ions are recorded with a time-to-digital converter (TDC). The number of counts as a function of laser frequency is our spectroscopy signal. In the 100~ms preceding the excitation, background counts (from ionization of background gas~\cite{Rengelink2018} and the scattered light from the dipole trap beams) are recorded. After exposure to the spectroscopy light, the atoms are released and fall under gravity onto a second MCP positioned below the optical dipole trap. By fitting the time-of-flight traces recorded with this MCP, we determine the chemical potential $\mu$ and the atom number of the BEC.
\begin{figure}[]
    \hspace{0.4cm} 
    \includegraphics[width=0.9\linewidth]{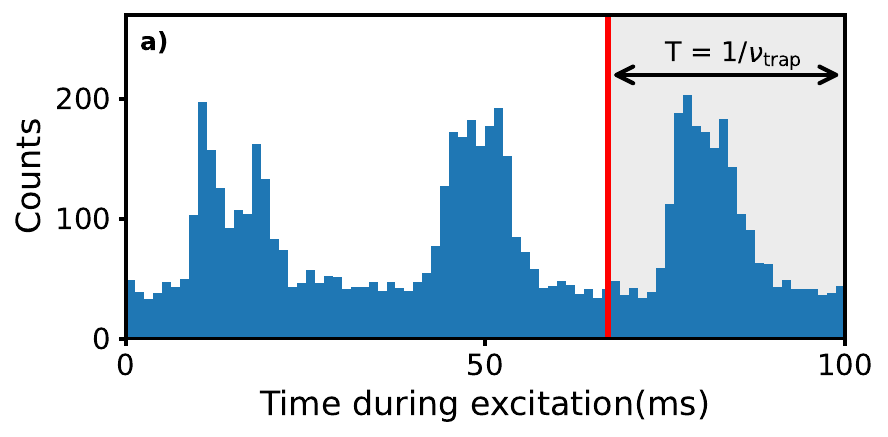}
    \includegraphics[width=0.9\linewidth]{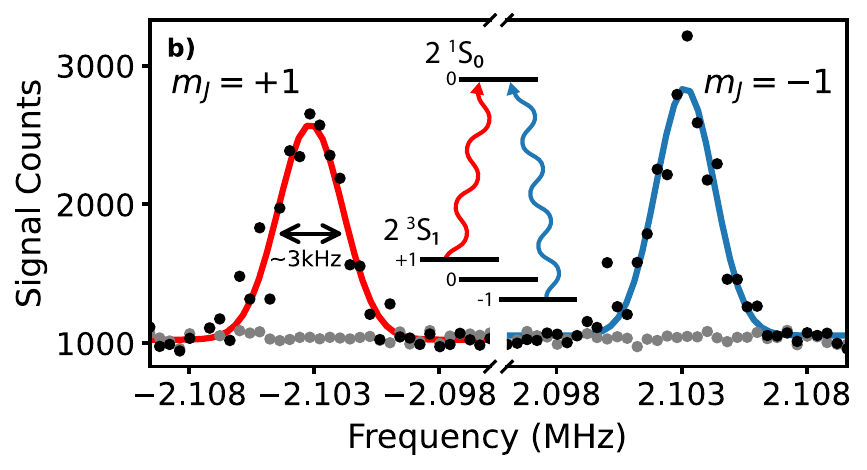}
    \caption{a) shows an histogram of from the arrival times of the He$^+$ ions over the 100~ms spectroscopy time measured with the time to digital converter. The axial trap oscillation induces a time-dependent Doppler shift periodic with the inverse of the trap frequency $\nu_{\text{trap}}$. As a result the BEC is only periodically resonant with the spectroscopy signal. To prevent a systematic Doppler shift, we sum the ion counts over a time window equal to the trap oscillation period, as indicated by the shaded area. Note that for the example shown here we deliberately increased the oscillation amplitude to illustrate the effect, but for all regular measurements the oscillations were first minimized (see text). b) shows a typical $m_J=\pm1$ spectrum. The $m_J = \pm1$ lines are alternately excited. The black points represent the ion counts with the spectroscopy laser on, while the grey points are background counts measured immediately before the spectroscopy. The typical transition linewidth is 3~kHz.}
    \label{fig:spectrum_example}
\end{figure}
To measure a spectrum, we repeat the above procedure for each spectroscopy laser frequency. We apply a magnetic field of 0.75~G, which is actively stabilized to cancel out changes in the ambient magnetic field between measurements. We alternate between exciting from the $2\,^3S_1$ $m_j =+1$ and the $m_j =-1$ states to the 2$\,^1S_0$ $m_J=0$ state and take the average to find the magnetic field-free transition frequency. Figure~\ref{fig:spectrum_example} shows an example of a measured spectrum. The observed linewidth of $\sim\!3$~kHz is attributed to a combination of mean-field broadening, magnetic noise at 50~Hz from the AC-line, the AC magnetic field generated by the turbomolecular pump used to evacuate the science chamber, and residual Doppler broadening from axial trap oscillations of the BEC.

Figure~\ref{fig:spectrum_example} shows an example of a histogram of the ion (signal) arrival times over a 100~ms spectroscopy exposure time. Peaks are observed when the velocity of the oscillating BEC induces a Doppler shift such that the atoms become resonant with the frequency of the spectroscopy laser. In \cite{Rengelink2018}, axial trap oscillations of the BEC, induced by beam pointing noise of the dipole trap beams, were identified as the dominant broadening mechanism. By increasing the mechanical stability of the setup, we have suppressed these oscillations due to pointing noise. However, we still observe oscillations induced by the switching of the magnetic trap and bias coils. Since the oscillations have a well-defined phase relationship with the experimental sequence, they can lead to a systematic frequency shift. We implemented two methods to cancel shifts. (i)  In post-processing, we set the spectroscopy exposure time equal to the axial trapping period, which is measured daily. This ensures that each measurement spans a complete oscillation cycle and that the net frequency shift averages to zero. (ii) We record two spectra, one for each of the two counter-propagating spectroscopy beam directions. This results in two data sets measured from opposite directions. By taking the average of both sides, we can suppress any residual Doppler shift.

The three systematic shifts that need to be extrapolated to zero are the ac Stark shift from the spectroscopy laser, the ac Stark shift from the dipole trap, and the mean-field shift arising from interactions between the atoms in the BEC. These frequency shifts scale linearly with the spectroscopy laser power, dipole trap power, and chemical potential, respectively. We measure these effects by varying the three parameters independently and fit a three-dimensional weighted linear regression model to the data to find the transition frequency. 

\begin{figure*}[]
    \centering
    \includegraphics[width=0.89\linewidth]{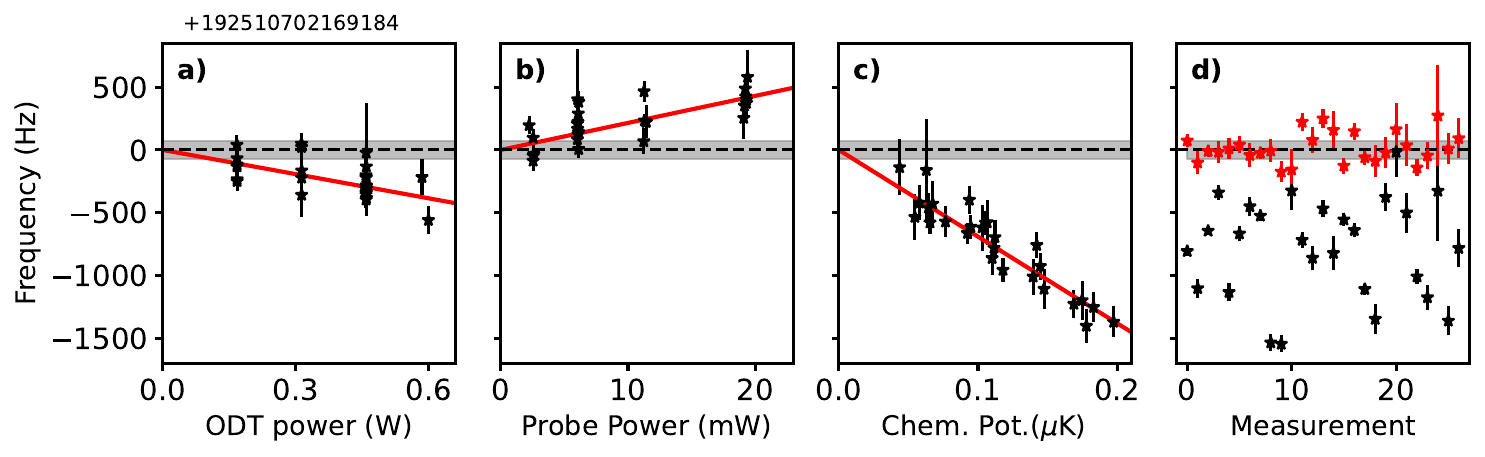}
    \includegraphics[width = 0.8\linewidth]{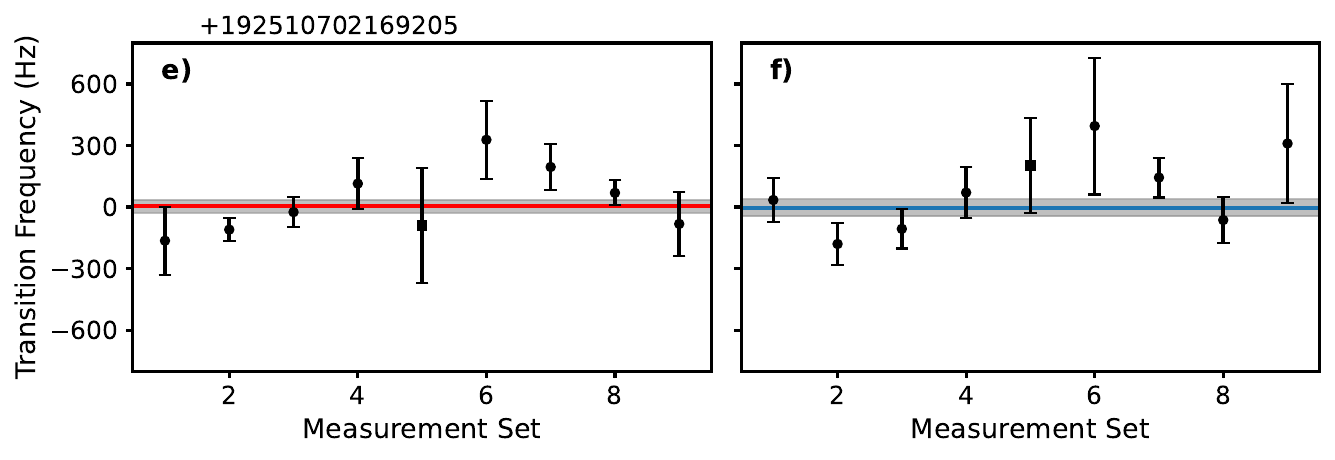}
    \caption{Overview of the multiple linear regression analysis and the resulting transition frequencies for all measurement sets. For measurement set 2, part a–c shows the partial residual plots as functions of the ODT power, probe power, and chemical potential, respectively. The slope of the ODT power regression yields the differential ac Stark shift, while the slope of the chemical potential regression is used to extract the triplet-singlet scattering length. The black circles in d) represent the measured frequencies for this data set, while the red points show the residuals after correcting for the measured systematic effects. Error bars correspond to the statistical uncertainty of the fit, with the residual errors bars adjusted to include the uncertainty from the fitted slopes. e) and f) show the measured transition frequencies from the various measurement sets with excitation from the left and right sides respectively. The solid lines represent the weighted mean, and the shaded areas indicate the standard error of the weighted mean. The result from the extrapolation shown in a–d corresponds to the transition frequency of set 3 in e). The square points represent measurements done with a local cesium clock.}
    \label{fig:extrapolation}
\end{figure*}

There is a 50\,ms delay between the end of the spectroscopy interaction and the release of the BEC from the trap. During this time, we turn off the bias coils, and if the atoms are in the $m_J = -1$ state, we transfer them to the $m_J = +1$ state for the measurement of the chemical potential. During this time, the chemical potential decays exponentially due to off-resonant scattering by the dipole trap light. We corrected the measured chemical potential for this decay. When all is taken into account, the extrapolation for a single measurement set  looks like the typical example shown in Figure~\ref{fig:extrapolation}a-d. 

The results of all nine measurement sets (based on 698 $m_J = \pm 1$ spectra) are shown in Figure~\ref{fig:extrapolation}e-f and were recorded between 28~November~2024 and 26~January~2025. The weighted averages of the two sides differ by \SI{9(53)}{Hz}. To further suppress Doppler shifts, we combine both sides by taking their arithmetic mean. We estimate that the residual Doppler shift is less than \SI{3}{Hz} (see Supplementary Material).
The measured transition frequency (see Table \ref{tab:systematic_shifts}) is affected by several systematic shifts and uncertainties that need to be taken into account. The most dominant, but easily calculable, correction arises from the recoil shift due to the absorption of a 1557\,nm photon. The next leading effect originates from magnetic field gradients caused by magnetic materials near the setup. If the $m_J = \pm 1$ states are spatially displaced within such a gradient, they experience different magnetic fields, and averaging the transition frequencies obtained from them does not completely cancel the Zeeman shift. To quantify this effect in our experiment, we compared the Zeeman shift experienced by the $m_J = \pm 1$ states by driving RF spin flips between both states, comparing the resonance conditions for atoms first prepared in $m_J = -1$ to those that started in $m_J = +1$. We found a difference of $-6(22)$~Hz, which is taken into account in Table~\ref{tab:systematic_shifts}. The uncertainty in the one-body lifetime, which we use to correct the measured chemical potentials, leads to an uncertainty of $33~\text{Hz}$ in the extrapolation to the unperturbed transition frequency; this includes the decay of the chemical potential during excitation. To check for any nonlinearities in the extrapolation of the chemical potential, we repeated the fit using only data points with chemical potentials greater than and less than \(0.11\,\mu\mathrm{K}\). No systematic deviation from the results presented here was found. Detailed derivations of all systematic shifts given in Table \ref{tab:systematic_shifts} can be found in the supplementary material \cite{remark1}. 
\begin{table}[]
    \centering
    \caption{Contributions to the transition frequency and uncertainty contributions (1$\sigma$
    ).}

    \label{tab:systematic_shifts}
    \resizebox{0.48\textwidth}{!}{  

    \begin{tabular}{ccc}
        \hline
        \hline
        Value/Correction & Value (Hz)  & Uncertainty (Hz) \\
        \hline
        Measured frequency & \SI{192510702169205.1}{}& 27 \\
        Recoil shift & \SI{-20554.30}{}  & 0.01 \\
        Magnetic field gradient& -6 & 22 \\
        Decay $\mu$ & 0& 33\\
        Residual Doppler shift & 0& 3\\

        Blackbody radiation & $+3.9$ & 0.3 \\
        DC-Stark shift ion MCP & 0& $0.6$ \\
        Clock  & 0 & $1$ \\
        
        Gravitational redshift  & $-0.09$ & 0.03  \\
        
        Quantum interference & $0$ & 0.1 \\
        Second-order Zeeman  & $<1\times10^{-3}$ & - \\
        Second-order Doppler  & $<1\times10^{-8}$ & - \\
        \hline
        Total & \SI{192510702148649}{} & 48 \\
        Rengelink \textit{et al.}\cite{Rengelink2018} & \SI{192510702148720}{}& 200 \\ 
        van Rooij \textit{et al.}\cite{vanRooij2011} & \SI{192 510 702 145.6}{}$\times10^3$& $1.8\times10^3$ \\ 
        
        \hline
        \hline
        
    \end{tabular}}
\end{table}
After applying all corrections, we find the transition frequency for the $2\,^3{S}_1\to2\,^1{S}_0$ to be \SI{192510702148649(48)}{Hz}. Our measurement is in excellent agreement with the previous best result by \cite{Rengelink2018}, but it is four times more precise. 

\begin{figure}[]
    \centering
    \includegraphics[width=0.9\linewidth]{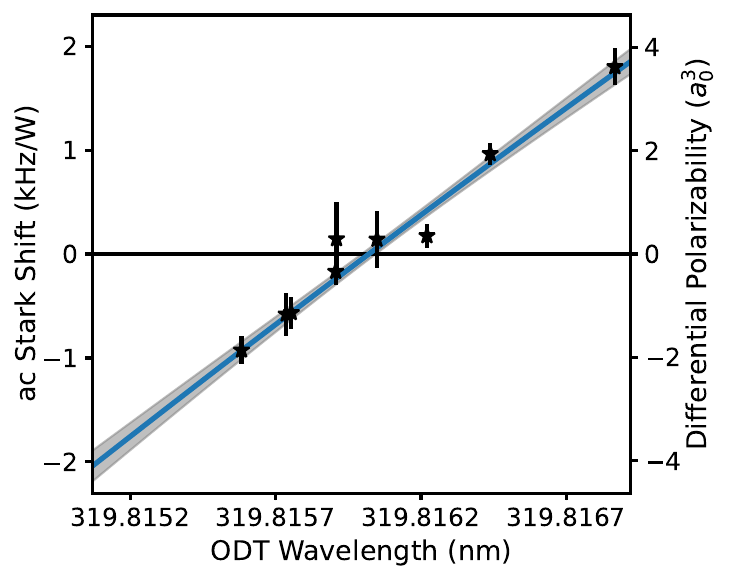}
    \caption{The measured (differential) ac Stark shift as a function of dipole trap wavelength between the $2\,^3{S}_1$ and $2\,^1{S}_0$ states. Each point represents the ac Stark shift determined from the multiple linear regression analysis for a specific measurement set. The blue curve is the fitted polarizability curve from \cite{Notermans2014}. The magic wavelength, found at the zero crossing of the differential ac Stark shift, is determined to be 319.81602(4)~nm. The grey shaded region indicates the 1$\sigma$ confidence interval on the fit.}
    \label{fig:magic_wavelength}
\end{figure}

For each measurement set, we changed the dipole trap wavelength and determined the corresponding differential ac Stark shift slope. To extract the nearest magic wavelength (where the differential ac Stark shift is zero), we use the (differential) polarizability curve from \cite{Notermans2014} to fit the measured slopes as a function of the optical dipole trap wavelength. This is shown in Figure~\ref{fig:magic_wavelength}. The magic wavelength we determined in this manner is equal to 319.81602(4)~nm. Our measurement agrees with the previous result reported by \cite{Rengelink2018}, but it is four times more precise. It also agrees within 2$\sigma$ with the most accurate theoretical prediction to date of \SI{319.8153(6)}{nm} from \cite{Wu_2018}, based on a relativistic full-configuration interaction calculation. 

Due to the difference in scattering lengths between the triplet and singlet states, atoms in the singlet state experience a different mean-field potential than those in the triplet state. The resulting frequency shift depends on the density. The inhomogeneous density distribution of the BEC, therefore, leads to broadening of the spectral line. However, due to other broadening mechanisms, this meanfield lineshape is not directly observable. We observe only the average shift as we vary the chemical potential. In terms of the $a_{tt}$ and $a_{ts}$ scattering lengths, this shift is given by \cite{Rengelink2018}:
\begin{equation}
    \braket{\Delta\nu_{meanfield}} \mu = -\frac{4}{7h}\bigg(\frac{a_{ts} - a_{tt}}{a_{tt}}\bigg)\mu
\end{equation}
 $\braket{\Delta\nu_{meanfield}}$ is the slope from the chemical potential regression. From the average of all measurement sets, including uncertainties for the decay of the chemical potential during spectroscopy, we find $\braket{\Delta\nu_{meanfield}} =  -6.74(30)~\text{kHz/}\mu\text{K}$. Using the measured value from~\cite{Moal2006} for the triplet-triplet scattering length $a_{tt} = +142.0(1)a_0$ we find $a_{ts} = +61.6(3.6)a_0$. This result differs by 3.3$\sigma$ from the result of \cite{Rengelink2018} $a_{ts} = +82.5(5.2)a_0$. We attribute this difference to an underestimation of the uncertainty in that work. Our measurement still differs significantly from the best calculation of $a_{ts} = 42^{+0.5}_{-2.5}a_0$ \cite{Notermans2016}, based on ab initio calculations of the involved molecular potentials \cite{muller1991}. Details regarding the systematic effects on the magic wavelength and the triplet–singlet scattering length are discussed in \cite{remark1}.

\begin{figure}[]
    \centering
    \includegraphics[width=0.99\linewidth]{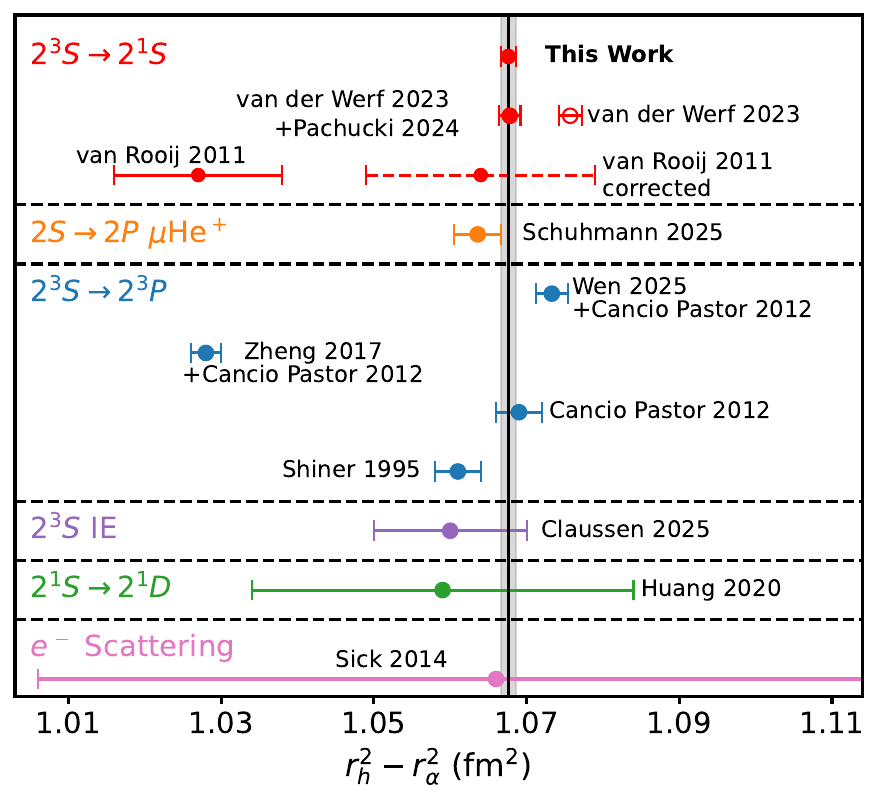}
    \caption{Overview of differential nuclear charge radius measurements between the alpha and helion particle, $r_{h}^2 -r_{\alpha}^2$. A comparison is made between measurements based on spectroscopy in electronic helium \cite{vanderWerf2025, vanRooij2011, Canciopastor2012, Canciopastor2004, Shiner1995,Zheng2017, Wen2025, ClausenMerkt2025, Huang2020}, muonic helium \cite{Schuhmann2025}, and electron scattering experiments \cite{Sick2014}. Zheng~\cite{Zheng2017} and Wen~\cite{Wen2025}  measured the $^{4}$He transition frequency and used the $^{3}$He result of Cancio Pastor~\cite{Canciopastor2012} to determine $r_{h}^2 - r_{\alpha}^2$.}
    \label{fig:DNCR_overview}
\end{figure}

Combining our result for the  $^4\text{He}$ $2\,^3{S}_1\to2\,^1{S}_0$ transition frequency with the measurement in  $^3\text{He}$ \cite{vanderWerf2025}, we can determine the isotope shift on this transition. To extract the charge radius difference (squared) from the isotope shift we use theory from \cite{Pachucki2024, pachucki2025qednuclearrecoileffect}, which has been improved to include the second order QED nuclear recoil correction, thereby reducing the theoretical uncertainty in the isotope shift calculation for a point nucleus to 107~Hz. Using that theory result and the procedure outlined in \cite{pachucki2025qednuclearrecoileffect}, we determine an improved value for the differential nuclear charge radius squared between the helion and the alpha particle of $r_{h}^{2} - r_{\alpha}^{2} = 1.0676(10)\,\text{fm}^{2}$. Our value is plotted in Figure \ref{fig:DNCR_overview}, along with  determinations from other experiments. Our value confirms our previous result \cite{vanderWerf2025, Pachucki2024}, but the uncertainty is 30\% smaller. 

In conclusion, we have measured the $2\,^3{S}_1 \to 2\,^1{S}_0$ transition in $^4\text{He}$ with 48 Hz accuracy, corresponding to a relative accuracy of $2.5\times10^{-13}$. This constitutes the most precise optical frequency measurement in helium to date. It was achieved with a novel method to suppress systematic Doppler shifts arising from BEC oscillations and by calibrating our spectroscopy laser to a remote active hydrogen maser via a White Rabbit fiber link.
Our improved determinations of the magic wavelength and the triplet–singlet scattering length provide stringent tests of the current theory of the helium atom.
The reported charge radius difference $r_{h}^2 -r_{\alpha}^2$ is the most accurate to date and confirms our previous value. It is in agreement within 1.2$\sigma$ with the value based on muonic helium ion spectroscopy~\cite{Schuhmann2025} and also within 0.8$\sigma$ with the one based on the ionization energy of the 2$\,^{3}S{_1}$ state.  This last aspect confirms that the isotope shift is not affected by the $\geq$9$\sigma$ discrepancy between the theoretical ionization energies of both states and the experimental values~\cite{ClausenMerkt2025}.

The accuracy of our $r_h^2 - r_\alpha^2$ determination is now dominated by the uncertainty in the theory of the isotope shift (107~Hz~\cite{pachucki2025qednuclearrecoileffect}) and the accuracy of 170~Hz from the previous $^3\text{He}$ spectroscopy measurement~\cite{vanderWerf2025}. The aim is to improve both to a similar accuracy as we have now demonstrated for $^4$He, which (providing sufficiently accurate theory will be available) would enable a determination of $r_{h}^2 -r_{\alpha}^2$ with an accuracy of $0.0004 \text{ fm}^2$.

\begin{acknowledgments}
We thank Joanne Slijper and Rob Kortekaas for technical support, Lex van der Gracht for the construction of the TDC, and Yuri van der Werf, Raphael Jannin, Frank Cozijn, Haoyuan Li, and Maarten Hoogerland for useful discussions. 
We acknowledge the support of SURF and VSL for providing access to UTC(VSL) via the SURF Time\&Frequency Network. K.E. acknowledges support from the Dutch Research Council (NWO) via grant 16MYSTP and 680-91-108.

\end{acknowledgments}

\bibliography{bibliography}

\end{document}


\maketitle

\section{Systematic effects on the determination of the transition frequency}
\subsection{Recoil shift}
The largest systematic shift is the recoil shift:
\begin{equation}
  f_{\mathrm{recoil}} = -\frac{h}{2 m \lambda^{2}}
  = \SI{-20554.30(1)}{\hertz},
\end{equation}
which arises from the absorption of a \SI{1557}{\nano\meter} photon.  The uncertainty in this value, dominated by the uncertainty in the helium mass~\cite{Prohaska2022, Mohr2025}, is negligible compared to the current level of experimental accuracy.

\subsection{Gradient of the magnetic field}
The cancelation of the Zeeman shift by averaging spectroscopy from the $2\,^3S_1\, m_J = \pm 1$ magnetic substates only holds if they experience the exact same magnetic field. Magnetic materials near the setup, such as the iron frame, distort the local field, producing a magnetic field gradient across the measurement region. Measuring the magnetic field at different positions along the direction of the spectroscopy laser, we determined a gradient of 0.28(0.14)~$\mu$G/mm or 0.80(40)~kHz/mm along this direction.

In the presence of such a magnetic field gradient, any spatial separation of the different magnetic substates can result in a systematic shift. To measure the magnetic field that atoms in the $m_J = \pm 1$ substates experience, we drive transitions between the two substates with an RF pulse and measure the population in the magnetic substates by conducting a Stern-Gerlach-type experiment: after the RF pulse, we let the cloud fall under gravity while applying a magnetic field gradient. In this way, we can spatially separate atoms in the different magnetic substates. Using absorption imaging, we measure the relative populations in the $m_J = \pm 1$ magnetic substates as a function of the applied RF frequency, from which the Zeeman splitting can be determined. We compare the measured Zeeman splitting when starting in the $m_J = +1$ and $m_J = -1$ states to verify whether both magnetic substates experience the same field. 
\begin{figure}[h]
    \centering
    \begin{minipage}{0.47\textwidth}
        \centering
        \includegraphics[width=\linewidth]{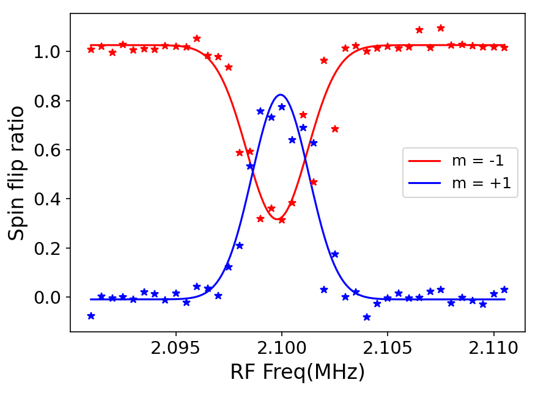}
        \caption{A Zeeman-splitting measurement is performed by interleaving spectra that start from the $m_J = +1$ and $m_J = -1$ states. From the difference in the fitted RF frequencies, we determine the difference in magnetic field experienced by atoms in the two magnetic substates.}
        \label{fig:spectrum}
    \end{minipage}
    \hfill
    \begin{minipage}{0.47\textwidth}
        \centering
        \includegraphics[width=\linewidth]{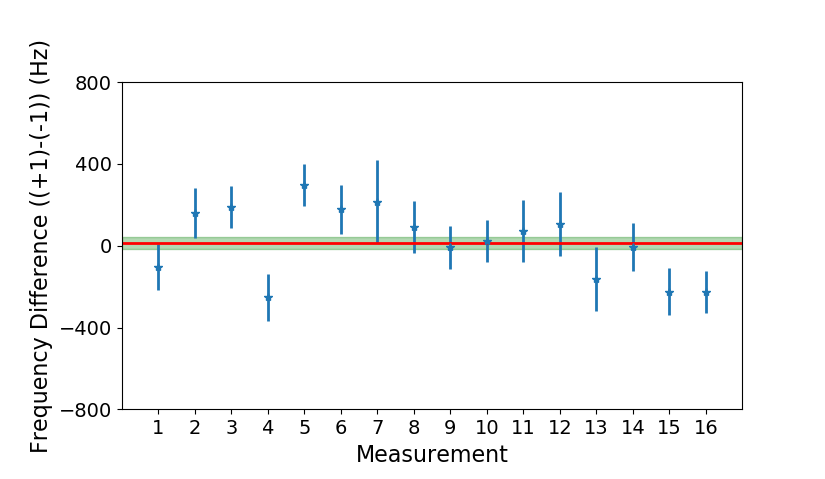}
        \caption{Measured differences in Zeeman splitting between atoms starting in the $m_J = +1$ and $m_J = -1$ states, with error bars representing the statistical uncertainty from the fits; the data were taken at a spectroscopy time of \SI{50}{ms} while the motorized waveplate was rotating. The red line shows the weighted average, and the green shaded area indicates the standard error of the mean.}
        \label{fig:overview_examples}
    \end{minipage}
    \label{fig:sidebyside}
\end{figure}
\begin{table}[h]
    \centering
    \begin{tabular}{|c|c|c|}
        \hline
        \textbf{Time} & \textbf{Waveplate} & \textbf{Difference in Zeemansplitting ((+1) - (-1))} \\ \hline
        50ms   & Rotating   & 14(30)Hz   \\ \hline
        63ms   & Rotating   & -36(33)Hz   \\ \hline 
        63ms   & Not Rotating  & -47(33)Hz   \\ \hline
    \end{tabular}
    \caption{Overview of measured differences }
    \label{tab:magnetic_field_differences}
\end{table}

Figure~\ref{fig:spectrum} shows an example of measured Zeeman splittings between the $m_J = +1$ and $m_J = -1$ states.  This measurement is repeated and averaged to decrease the statistical uncertainty. An overview of a measurement set is shown in Figure~\ref{fig:overview_examples}. The experimental sequence is identical to that used during optical spectroscopy, except that, instead of the 100~ms IR spectroscopy duraction, a \SI{200}{\micro\second} RF pulse is applied. Since the RF pulse is much shorter than the optical spectroscopy time, it does not probe a full trap oscillation period. To investigate the motion of the atoms in the $m_J = +1$ and $m_J = -1$ states within the trap, which may influence the magnetic field they experience, the difference in Zeeman splitting is measured at various times during the sequence.

During optical spectroscopy, the primary differences between addressing the two states are the RF sweep, which induces Landau-Zener transitions from $m_J = +1$ to $m_J = -1$, and the polarization of the spectroscopy laser, which is controlled by a motorized rotation stage. The rotation stage contains permanent magnets that alter the magnetic field in the experiment. While the waveplate itself is located far from the atoms and does not directly affect them, it is positioned near a magnetic field sensor (\SI{20}{cm} away) that provides feedback to the atoms. To investigate whether the motorized waveplate affects the Zeeman shift, we compare measurements taken with and without the rotation of the waveplate.

The results of these measurements are presented in Table~\ref{tab:magnetic_field_differences}; the measurement with the waveplate stationary differs from those taken while it was rotating. However, we observe no correlation between the waveplate position and the magnetic field at the sensor itself (at the $\mu$G level); therefore, we attribute this difference to a statistical fluctuation. 

The measurements at different times (when the waveplate is rotating) are consistent with a no shift within the errorbars. However, as the BEC is oscillating, there is a gradient in the magnetic field during the oscillation;  The measured magnetic field at this time could still differ from that during the RF spectroscopy, as the relevant timescales of the two measurements are different. 

The maximum change in position can be estimated from the oscillation amplitude: a typical maximum Doppler shift of $500~\text{Hz}$ at a trapping frequency of $25~\text{Hz}$ corresponds to an oscillation amplitude of approximately $5~\mu\text{m}$. Hence, the maximum difference compared to the center of the trap is about $5~\mu\text{m}$. Assuming a conservative magnetic-field gradient of $1200~\text{Hz/mm}$ (corresponding to the measured slope plus one standard deviation), this results in a maximum frequency error of approximately $6~\text{Hz}$.

To correct our measurement, we take the average of the two values obtained while the waveplate was rotating, as this most accurately represents the actual spectroscopic conditions. This yields an average of \(-11(32)\,\mathrm{Hz}\), where the uncertainty corresponds to the mean of the two individual error bars. Since the spectroscopy involves averaging the \(\pm1\) transitions, the mean value is divided by two. To account for the oscillating condensate (BEC) during the magnetic field measurement, we include an additional systematic uncertainty of \(\pm6\,\mathrm{Hz}\), which we add linearly.  The resulting correction to the transition frequency is $-6(22)\mathrm{Hz}$.

\subsection{Decaying chemical potential}
Because of off-resonant scattering from the UV light of the dipole trap, the number of atoms in the trap, and hence the chemical potential, decreases over time. There is a \SI{50}{ms} delay between the end of the spectroscopy and the release of the BEC from the trap, as the $m_J = -1$ atoms are transferred back to the $m_J = +1$ state to measure the chemical potential. As a result, the measured chemical potential is lower than that experienced by the atoms at the end of the spectroscopy period. As the loss rate depends exponentially on the UV power, this may lead to errors when we linearly extrapolate to find the transition frequency.

The exponential decay of the chemical potential is given by \cite{Notermans2016}:
\begin{equation}
    \mu(t) = \mu_0e^{-\frac{2}{5}t\Gamma}.
    \label{eq:decay}
\end{equation}
Here, $\Gamma$ is the scattering rate, i.e., the inverse of the one-body lifetime $\Gamma = 1/\tau$, and it scales linearly with the UV power. To correct for this effect, we multiply the measured chemical potential by the inverse of equation~\ref{eq:decay} evaluated at $t = \SI{50}{ms}$, using measured lifetimes ($\tau \approx 0.5$--$2\,\text{s}$) that depend on the UV power.

We have not measured the BEC lifetimes routinely. Since the lifetime depends on the intensity of the dipole trap, which may vary with the alignment of the two beams, the lifetime also changes over the course of the measurement. To estimate the variation in lifetime, we examined the variation in trap frequency at a given power over the measurement set. As the trap frequency is proportional to the square root of the peak intensity, the variation of the lifetime should be proportional to the squared variation of the trap frequency. Based on the worst- and best-case alignments, we estimate conservatively that the variation in intensity, and hence the scattering rate, is less than 50\%. To assess the resulting uncertainty in frequency, we vary the scattering rate by 50\% and perform the extrapolation, yielding a frequency deviation of $\Delta f = \pm 13~\text{Hz}$. 

The chemical potential also decays during the spectroscopy itself, not only due to off-resonant scattering but also because of excitation to the singlet state. Consequently, the chemical potential at the end of the spectroscopy period does not correspond to its average value over the entire duration of the spectroscopy. Although this difference is small, since the spectroscopy time is short compared to the lifetime of the cloud, it can nonetheless introduce a slight nonlinearity in the regression toward zero, leading to a small frequency shift. 

To quantify the magnitude of the frequency shift introduced in the extrapolation, we simulated the spectroscopy. The decay of the chemical potential over time is governed by \cite{Notermans2016}:
\begin{equation}
    \frac{d\mu}{dt} = - \frac{2}{5}\Gamma\mu - \frac{2}{5}\mu\, S(\mu, \nu(t), U_0, P_{\text{spectro}}).
    \label{eq:mu_decay}
\end{equation}
Here, $S(\mu, \dots)$ represents the mean-field lineshape. 
The laser frequency in the frame of the atoms, $\nu(t)$, includes the oscillating Doppler shift as well as the influence of 50~Hz and 833~Hz ac magnetic fields. $U_0$ denotes the trap depth, and $P_{\mathrm{spectro}}$ denotes the spectroscopy power. The spectroscopy power determines the depletion resulting from atoms being pumped to the singlet state and is chosen to match the experimentally observed depletion (maximally $10\%$ at maximum spectroscopy power). It is not possible to solve this equation analytically; hence, we must solve it numerically. Using the calculated time-dependent chemical potential, we calculate and fit a spectrum, and repeat this procedure for various chemical potentials, trap depths, and spectroscopy laser powers, followed by a regression to find the extrapolated transition frequency. The effect of the decay can be quantified by evaluating how the transition frequency and slopes deviate from their initial values at the start of the simulation.

When only the decay of the chemical potential due to excitation is taken into account, we find a frequency deviation of $-12~\mathrm{Hz}$. When only off-resonant scattering is included and the decay due to excitation is neglected, the deviation is $+16~\mathrm{Hz}$. If we increase the scattering rate by $50\%$ (which corresponds to the maximum variation of the off-resonance we expect during the measurement), this deviation increases to $20~\mathrm{Hz}$. When both effects are included simultaneously (using the measured scattering rate), the net deviation is $0~\mathrm{Hz}$. Thus, the two effects cancel each other in the extrapolation, and no correction is applied to the measured transition frequency.

However, depending on the precise off-resonant scattering rate, the shift can range from $-12~\mathrm{Hz}$ to $20~\mathrm{Hz}$. We therefore take a conservative value of $20~\mathrm{Hz}$ for the uncertainty associated with this effect. This uncertainty is added linearly (as both effects depend on the uncertainty/variation in the one-body lifetime) to the uncertainty due to the $50~\mathrm{ms}$ decay observed at the end of the spectroscopy. The combined uncertainty is therefore $33~\mathrm{Hz}$.

Since the decay of the chemical potential affects the slope of both the mean-field shift and AC Stark shift regressions, it also influences the determination of the magic wavelength and the singlet-triplet scattering length. Corrections are provided below.


\subsection{Residual Doppler shift}
If the spectroscopy duration matches the oscillation period of the trapped atoms, the net Doppler shift averages to zero. However, since the trapping frequency is measured only once per day, it can drift during the measurements due to slow changes in beam alignment, introducing a residual Doppler shift. To mitigate this, we perform measurements from opposite of the trap, creating two interleaved data sets that are then averaged to determine the transition frequency. Because the measurements are not simultaneous, drifts in trapping frequency between them can still produce a small systematic difference in Doppler shift. By monitoring the trapping frequency over a 10-hour measurement day, we found that it can drift up to 10\% between the start and the end of the measurement day. Measurements from opposite directions are always taken 15 minutes apart. 

To estimate the residual Doppler shift due to trapping frequency drift, we perform a simple calculation: we consider the maximum observed oscillation amplitude (1~kHz) and assume a 10\% drift while keeping the spectroscopy time fixed at the initial trapping frequency. The oscillation phase is chosen to maximize the residual Doppler shift. Between measurements from opposite directions, we assume a 0.25\% difference in trapping frequency (based on 10 \% drift over 10 hours). Averaging over the two directions gives a  residual Doppler shift of 3~Hz, which we take as an uncertainty estimate.  Furthermore, the oscillation phase and exact trapping frequency vary randomly between measurements, so the residual Doppler shift largely averages out over the entire data set.

\subsection{Nonlinear effects in the chemical potential extrapolation}
To check for possible nonlinearities in the extrapolation of the chemical potential (the largest systematic effect that we extrapolate), we repeated the fit using only data points with chemical potentials greater than $0.11~\mu\mathrm{K}$. This yielded an extrapolated value of \SI{192510702169211(43)}{Hz}, in full agreement with the measurement presented in the main text. The mean-field slope obtained from this fit, $-6.47(29)~\mathrm{kHz}/\mu\mathrm{K}$, is also consistent with the previously reported value. 

Including data points with chemical potentials smaller than $0.11~\mu\mathrm{K}$ likewise yields results consistent with those reported in the paper: the extrapolated transition frequency is\\
\SI{192510702169087(130)}{Hz}, and the corresponding mean-field slope is $-5.8(1.1)~\mathrm{kHz}/\mu\mathrm{K}$.

\subsection{Frequency reference}
\label{subsec:clock}
UTC(VSL)—the local realization of UTC in the Netherlands, maintained by VSL in Delft—is disseminated to our laboratory in Amsterdam through an optical fiber using the White Rabbit protocol. 
The long-term fractional frequency deviation of UTC(VSL) with respect to UTC is specified to be less than $5\times10^{-15}$ fractionally \cite{private_communication}. Indeed, inspection of the CircularT data reported for the period during our measurement campaign indicates a fractional frequency stability of the deviations y(UTC-UTC(VSL)) of $3\times10^{-15}$ for an 8-day averaging time \cite{circularT}. The traceability to the SI second of UTC(VSL) furthermore receives an uncertainty contribution of $1.2\times10^{-16}$  from the fractional difference, $d_{TAI}$, between TAI and Terrestrial Time, where the latter represents the coordinate time that is defined in terms of the SI second, valid for observers on the geoid \cite{Petit2018}. We conservatively take the $5\times10^{-15}$ frequency deviation specified for the WR link to UTC(VSL) as the uncertainty of the representation of the SI second in our laboratory.


The transition frequency measured in set 5 for both excitation directions was referenced to a local cesium clock. These measurements were corrected for the offset of this cesium clock relative to the SI second, determined by comparing the pulse-per-second output of the clock with a GPS receiver over multiple days. The resulting correction applied to the transition frequency for these sets was 
-35~Hz. We take the noise floor of the cesium clock of $5\times10^{-14}$ as the uncertainty, which corresponds to 10~Hz on our transition. 

\subsection{DC-Stark shift ion MCP}
The difference in dc polarizability between the singlet and triplet states is 483.55 \(\times 4\pi \ \epsilon_0\,  a_0^3\) \cite{Notermans2014}. The electric field generated by the ion MCP at the position of the BEC induces a dc Stark shift:
\begin{equation}
\Delta\nu = -\frac{1}{2h}\text{Re}(\Delta\alpha)E^2 .
\label{eq:DC_STARK_SHIFT}
\end{equation}
The estimated electric field at the position of the atoms is 30 V/m. To account for potential uncertainties, we scale this value by a factor of 10, which leads to a conservative estimate of the DC Stark shift of $\Delta \nu_{MCP} \approx$ $-0.6$~Hz \cite{Rengelink2018}. Since we do not know the exact electric field, we do not correct the measurement with this value; instead, we add it to the uncertainty.  

\subsection{DC-Stark shift blackbody radiation}
The typical temperature in the lab is 295~K. The mean squared electric field from blackbody radiation at 295~K is given by 
\(\langle E^2 \rangle = (804~\text{V}\,\text{m}^{-1})^2\)~\cite{Rengelink2018}. This means the transition is shifted by  \(\Delta \nu_{BB} = -3.9~\text{Hz}\). The maximum temperature variation over the whole measurement campaign in the lab was \(\pm 5~\text{K}\), and we use that to estimate the uncertainty of the shift. This corresponds to a frequency uncertainty of 0.3~Hz. Therefore, we correct the measured transition frequency by +3.9(3)~Hz
 
\subsection{Gravitational redshift}
As discussed in section \ref{subsec:clock}, our reference clock is located at VSL, the national metrology institute, in a different city approximately 50~km from our laboratory. The time scale UTC(VSL) serves as the reference for the White Rabbit link, which implies that a White Rabbit terminal located at the geoid (sea level) can produce an SI-traceable frequency with zero  gravitational redshift (see also Sec. 1.6). The ground level of our laboratory is roughly 0.8~m below sea level \cite{AHNViewer}, and the laboratory itself is located about 4~m above ground level (first floor). With the atoms sitting 1 meter above the floor. We therefore estimate the total height difference between the geoid and our laboratory as $4(1)$~m. This corresponds to a relative frequency shift of $+4.4(1.1)\times10^{-16}$, which translates to a correction of $-0.09(3)$~Hz to the measured transition frequency.


\subsection{Quantum interference shift}
Off-resonant excitation of distant transitions can lead to a frequency shift in spectroscopy; an effect known as \emph{quantum interference}. In \cite{Rengelink2018}, this effect on the $2\,^3S_{1} \rightarrow 2\,^1\!S_{0}$ transition was estimated to be less than $100\ \mathrm{mHz}$. We include this contribution as part of the uncertainty in our measured transition frequency.

\subsection{Second-order Zeeman shift}
The first-order Zeeman shift in the $2\,^3S_1$ state is canceled out by alternating between measurements from the $m_j = +1$ and $m_j = -1$ magnetic substates. The $2\,^1S_0$ state does not have a first-order Zeeman shift. The second-order Zeeman shift resulting from coupling to higher electronic states does not cancel in our experimental procedure. For the  $2\,^3S_1$ and $2\,^1S_0$ states, the second-order Zeeman shift is 2.3 mHz/G$^2$ and 3.2 mHz/G$^2$, respectively \cite{Vassen2018}. During our spectroscopy measurements, a magnetic field of 0.75~G is applied. The corresponding correction to the transition frequency is smaller than 1~mHz. This effect is completely negligible at our experimental accuracy.

\subsection{Line pulling from excitation of the thermal cloud}
Besides the BEC, there is also a thermal fraction present in the optical dipole trap. Based on a temperature of 110~nK, we expect a linewidth of about 30 kHz for this thermal cloud, which is much larger than the typical spectral linewidth of the BEC. Since the BEC resonance is slightly shifted due to mean-field interactions, the broad thermal component can, in principle, pull the measured line. We did not observe any thermal component beneath our measurements. Furthermore, simulations show that even for a relatively large thermal fraction of 30\% at a temperature of 110 nK, the resulting line shift is only on the order of a few hertz. However, in the extrapolation of the chemical potential, this contribution is extrapolated to zero, leading to no observable shift.

\subsection{Second-order Doppler shift}
The BEC oscillates in the optical dipole trap with a time dependent velocity $v(t)$. When the atoms are excited by the spectroscopy laser ($\lambda = 1557$ nm), the frequency observed in the frame of the atoms is shifted. The  fractional frequency shift $\Delta f(t)/f $, is given by \cite{Fisk_1997}:
\begin{equation}
    \frac{\Delta f(t)}{f} = \frac{v(t)}{c} - \frac{1}{2}\Big(\frac{v(t)}{c}\Big)^2+\frac{1}{2}\Big(\frac{v(t)}{c}\Big)^3 - \dots
\end{equation}
The first term in the expansion corresponds to the first-order Doppler effect. This is canceled by excitation from opposite directions and by restricting the excitation time to the inverse of the trapping frequency. The second term in the expansion is the second-order Doppler shift, given by $\Delta f_{\rm 2OD}/f = -\frac{1}{2}(v/c)^2$. To estimate the maximum possible shift, we use the maximum measured Doppler shift, $\Delta f = 2~\mathrm{kHz}$, which corresponds to a velocity $v = \lambda \Delta f \approx 3~\mathrm{mm/s}$. For this velocity, the fractional second-order Doppler shift is $\Delta f_{\rm 2OD}/f \approx 5 \times 10^{-23}$, or in frequency units, $\Delta f_{\rm 2OD} \approx -10~\mathrm{nHz}$. This effect is therefore negligible at the current level of experimental accuracy.

\section{Systematic effects on the determination of the magic wavelength}
From the fit on the ac Stark shift as a function of the wavelength of the laser that creates the optical dipole trap, we find a magic wavelength of 319.81596~nm, with the uncertainty from the fit itself of $2.5\times10^{-5}$~nm. Several systematic effects contribute to the uncertainty in the determination of the magic wavelength. First of all, the uncertainty of the lifetime due to off-resonant scattering also affects the magic wavelength. 

We again differentiate between two effects: the decay during the spectroscopy and the decay occurring during the final $50~\mathrm{ms}$ after the spectroscopy. For the latter, assuming the same $50\%$ variation of the scattering rate (as used for the spectroscopy to correct the chemical potential during the last $50~\mathrm{ms}$ of the sequence), we find a variation of $1.6 \times 10^{-5}~\mathrm{nm}$ in the magic wavelength. We therefore take this as the uncertainty induced by this effect. 

In the simulation used to quantify the effect of the decay of the chemical potential during spectroscopy, we found a shift of the magic wavelength of $1.0 \times 10^{-5}~\mathrm{nm}$ when the off-resonant scattering rate was increased by $50\%$. We use this as a conservative estimate of the uncertainty of this effect. Combined linearly with the uncertainty induced by the decay of the chemical potential during the last $50~\mathrm{ms}$ (as the two are correlated), we find a total uncertainty of $2.6 \times 10^{-5}~\mathrm{nm}$.

Secondly, the frequency of the 1085~nm laser is locked to a wavemeter. The wavemeter is regularly calibrated to a He-Ne laser for short term drifts, and we calibrate the absolute frequency with a laser locked to a helium cell on the $2\,^3{S}_1\rightarrow2\,^3{P}_2$ transition. From our calibration measurements, we find an offset of -90~MHz with a variation of $\pm$ 5~MHz between measurements. This leads to a correction of $+6\times10^{-5}$~nm with an uncertainty of $0.4\times10^{-6}$~nm on the magic wavelength. These effects are summarized in table \ref{tab:magic_wl}.  The effects of tensor and vector polarizability are negligible \cite{Rengelink2018}.
\begin{table}[h!]
    \centering
    \caption{Contributions to the magic wavelength}

    \label{tab:magic_wl}

    \begin{tabular}{ccc}
        \hline
        \hline
        Value/Correction & Value (nm)  & Error (nm) \\
        \hline
        Measured magic wavelength & 319.81596& $2.5\times10^{-5}$ \\
        Decaying chemical potential & 0  &  $2.6\times10^{-5}$ \\
        Wavemeter Correction& $+6\times10^{-5}$ & $0.4\times10^{-5}$ \\ 
        \hline
        \hline
        Total &  319.81602 & $4\times10^{-5}$

    \end{tabular}
\end{table}
\section{Systematic effects on the mean-field  slope}
From the weighted average of the slopes fitted to the transition frequency as a function of the chemical potential for all individual data sets, we find an average slope of $-6.74(16)\text{ kHz/}\mu\text{K}$. The uncertainty of the lifetime, due to off-resonant scattering, which we use to correct the chemical potential for the last 50~ms of the sequence, also affects the triplet-singlet scattering length. Based on the same 50\% variation in the scattering rate as considered previously, we find a variation of $\pm 0.1~\text{kHz}/\mu\text{K}$ in the measured slope of the mean-field shift. In the simulation we performed to assess the effect of the decaying chemical potential during the $100~\mathrm{ms}$ spectroscopy, we did not find any influence of the decay on the mean-field slope. However, when the scattering rate is increased by 50\%, the slope increases by $0.15~\text{kHz}$. We therefore take $\pm 0.15~\text{kHz}$ as a conservative estimate of the uncertainty in the slope due to decay during excitation. This uncertainty is added linearly to the shift arising from the uncertainty in the final $50~\text{ms}$ of the sequence. Combining all these effects in quadrature yields a mean-field shift slope of $-6.74(30)~\text{kHz}/\mu\text{K}$.

\section{Calculation of the differential nuclear size squared}
To calculate the differential squared nuclear charge radius between the alpha and the helion particle, we use the theoretical values for the isotope shift on the $2\,^3S_1 \to 2\,^1S_0$ transition for nuclei, together with the conversion factors from \cite{pachucki2025qednuclearrecoileffect}. Details are provided in table \ref{tab:nuclear_radius}.
\begin{table}[h!]
\centering
\caption{Contributions to the nuclear charge radius difference $\delta r^2 = r_h^2-r^2_{\alpha}$ and related quantities. }
\label{tab:nuclear_radius}
\resizebox{0.99\textwidth}{!}{  

\begin{tabular}{l r l}
\hline\hline
Contribution & Value (kHz) & Reference \\
\hline
$E({}^3\mathrm{He},\, 2\,^3S_{F=3/2}\to2\,^1S_{F=1/2} )$ & \SI{192 504 914 418.960(170)}{} & Exp.~\cite{vanderWerf2025} \\
$-E({}^4\mathrm{He,} \, 2\,^3S \to 2\,^1S)$ & \SI{-192\,510\,702\,148.649(48)}{} & This work \\

$\delta E_{\mathrm{hfs}}(2\,^3S_{F=3/2})$ & $-2\,246\,567.059(5)$ & Exp.~\cite{Rosner1970, Schuessler1969} \\
$-\delta E_{\mathrm{iso}}(2\,^3S \to 2\,^1S)$ (QED, point nucleus) & 8\,034\,146.901(105) & Theory \cite{pachucki2025qednuclearrecoileffect} \\
$-\delta E_{\mathrm{iso}}(2\,^3S \to 2\,^1S)$ (hyperfine mixing) & -79.056 & Theory \cite{pachucki2025qednuclearrecoileffect} \\
$-\delta E_{\mathrm{iso}}(2\,^3S \to 2\,^1S)$ (nuclear structure) & 0.157(20) & Theory \cite{pachucki2025qednuclearrecoileffect} \\
\hline
Sum ($E_{fs}$) &$-228.746(207)$ & \\
$C$ & $-214.758$ kHz/fm$^2$ &  Theory \cite{pachucki2025qednuclearrecoileffect}\\
$D$ & 0.017 kHz/fm$^2$ &  Theory \cite{pachucki2025qednuclearrecoileffect}\\
\hline
$\delta r^2 = r_h^2-r^2_{\alpha}$ & 1.0676(10) fm$^2$ & This work \\
\hline\hline
\end{tabular}
}
\end{table}
\\
The finite size effect ($E_{fs}$) is related to the square of the nuclear size difference as:
\begin{equation}
   E_{fs}[{}^{3}\text{He} - {}^{4}\text{He}] \;\equiv\; 
C \,\big( r_h^2-r^2_{\alpha}\big) \;+\; D \,\big( r_h^2+r^2_{\alpha}\big),
\end{equation}
Note that, in contrast to earlier modeling of the charge radius difference squared \cite{Pachucki2015}, the sum of the charge radii now also plays a (minor) role through parameter $D$, requiring knowledge of the absolute charge radii.

\bibliographystyle{unsrt}  
\bibliography{bibliography_sup}   